\documentclass[english,aps,prl,twocolumn,groupedaddress,showpacs]{revtex4}
\usepackage[T1]{fontenc}
\usepackage[latin1]{inputenc}
\usepackage{amsmath}
\usepackage{graphicx}
\usepackage{amssymb}

\makeatletter

\makeatletter

\usepackage{graphics}

\usepackage{epsfig}

\usepackage{epsfig}

\usepackage{amsfonts}

\newcommand{\BigR}{{\rm I\kern-.27em R}}
\newcommand{\BigN}{{\rm I\kern-.25em N}}

\newcommand{\Sum}{\displaystyle {\sum}}
\newcommand{\be}{\begin{equation}}
\newcommand{\de}{\end{equation}}

\makeatother

\usepackage{babel}
\makeatother
\begin{document}

\title{Random changes of flow topology in two dimensional and geophysical
turbulence}

\author{Freddy Bouchet}

\email[]{Freddy.Bouchet@inln.cnrs.fr}

\homepage[]{www.inln.cnrs.fr/rubrique.php3?idrubrique=79}

\author{Eric Simonnet}

\affiliation{INLN, CNRS, UNSA, 1361 route des lucioles, 06 560 Valbonne, France}

\pacs{47.27.-i, 47.27.E-, 92.60.hk, 05.40.Ca}

\keywords{Large scales of turbulent flows, 2D turbulence, geophysical turbulence,
out of equilibrium statistical mechanics, bistability.}

\begin{abstract}
We study the two dimensional (2D) stochastic Navier Stokes (SNS) equations
in the inertial limit of weak forcing and dissipation. The stationary
measure is concentrated close to steady solutions of the 2D Euler
equation. For such inertial flows, we prove that bifurcations in the
flow topology occur either by changing the domain shape, the nonlinearity
of the vorticity-stream function relation, or the energy. Associated
to this, we observe in SNS bistable behavior with random changes from
dipoles to unidirectional flows. The theoretical explanation being
very general, we infer the existence of similar phenomena in experiments
and in models of geophysical flows. 
\end{abstract}
\maketitle
The largest scales of turbulent flows are at the heart of a number
of geophysical processes : climate, meteorology, ocean dynamics, the
Earth magnetic field. The Earth is affected on a very large range
of time scales, up to millennia, by the structure and variability
of these flows. Many of these undergo extreme and abrupt qualitative
changes, seemingly randomly, after very long period of apparent stability.
This occurs for instance for magnetic field reversal for the Earth
or in MHD experiments \cite{Berhanu_etc_Fauve_2007_EPL_MagneticFieldReversal},
for 3D flows \cite{Ravelet_Marie_Chiffaudel_Daviaud_PRL2004}, for
multiple equilibria of atmospheric flows \cite{Weeks_Tian_etc_Swinney_Ghil_Science_1997},
for 2D turbulence experiments \cite{Sommeria_1986_JFM_2Dinverscascade_MHD,Maassen_Clercx_VanHeijst_2003JFM}
and for the paths of the Kuroshio and Gulf Stream currents \cite{Schmeits_Dijkstraa_2001_JPO_BimodaliteGulfStream}. 

Understanding these phenomena requires a statistical description of
the largest scales of turbulent flows. Very few theoretical approaches
exists due to the prohibitively huge number of degrees of freedom
involved. Fruitful hints may be drawn from qualitative analogies with
bistability in system with few degrees of freedom perturbed by noise
\cite{Benzi_2005PRL}. However the range of validity of this approach
remains a tough scientific issue, because of the complexity of turbulent
flows. What is the good theoretical framework for such phenomena ?
In the following, we argue that 2D turbulence, because of its relative
theoretical simplicity, is a very interesting framework in order to
address such an issue.

\begin{figure}[htpb]
\includegraphics[width=1\columnwidth,keepaspectratio]{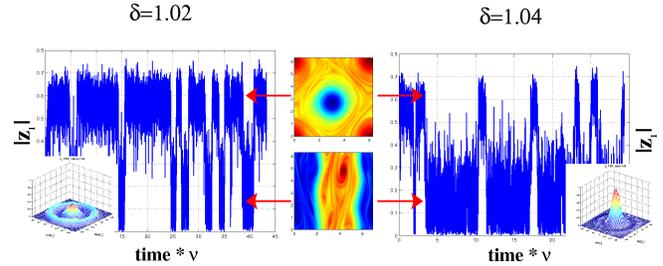}

\caption{Time series and probability density functions (PDFs) for the order
parameter $z_{1}$ (see page 4) illustrating random changes between
dipoles and unidirectional flows.\label{ts_f}}
\end{figure}

In this letter we predict and prove the existence of random switches
from dipoles to unidirectional flows (see Fig \ref{ts_f}), in the
2D Navier Stokes Eq. with random force (SNS). Similar random changes
have already been observed in rotating tank experiments for quasi-geostrophic
dynamics \cite{Weeks_Tian_etc_Swinney_Ghil_Science_1997}. Following
analogous theoretical considerations as for SNS Eq., we infer that
such changes will generically occur within a large class of models
like quasi-geostrophic (QG) or shallow-water (SW) models that describe
atmospheric \cite{Weeks_Tian_etc_Swinney_Ghil_Science_1997}, and
oceanic \cite{Schmeits_Dijkstraa_2001_JPO_BimodaliteGulfStream} large
scales. The recipe we propose is to exhibit bifurcation lines representing
abrupt change in steady solutions in the inertial limit and then look
for the corresponding transitions in real flows.

Geophysical and 2D inviscid flows are characterized by the conservation
of energy and an infinite number of quantities (Casimirs), such as
enstrophy. This property prevents direct energy cascade towards the
small scales, by contrast with 3D turbulence. Then, the first phenomenon
is an inverse energy cascade towards the large scales and a direct
enstrophy cascade. Kraichnan classical theory \cite{Kraichnan_Phys_Fluid_1967_2Dturbulence}
studies the self-similar processes associated with these two cascades
(see the recent spectacular discovery of conformal invariance consequences
for the inverse cascade \cite{Bernard_Boffetta_Celani_Falkovich_2006Nature}).
The second phenomenon, the self organization of the flow into jets
and vortices, occurs if energy is not dissipated before reaching the
largest scale. Then coherent structures break the self-similarity
so that their study cannot be properly addressed using Kraichnan theory
\cite{Kraichnan_Phys_Fluid_1967_2Dturbulence}. A second classical
theory, the so-called Robert-Sommeria-Miller (RSM) equilibrium statistical
mechanics \cite{Eyink_Sreenivasan_2006_Rev_Modern_Physics}, predicts
the self-organized structures for inviscid decaying turbulence. However,
this inviscid theory does not take into account the long-term effects
of forcing and dissipation as well as the slow dynamics of the flow.
Therefore, random changes of flow topologies cannot be explained by
these two classical theories.

As an alternative theoretical approach, we study statistically stationary
states of SNS Eq. Note that a self-similar growth of a dipole has
been studied in \cite{Chertkov_Connaughton_andco_2007_PRL_EnergyCondesation}
emphasizing transient growths: both approaches complement each other.
SNS Eq. on a doubly-periodic domain $\mathcal{D}=(0;2\pi\delta)\times(0;2\pi)$
reads {\small \begin{equation}
\frac{\partial\omega}{\partial t}+\mathbf{v}\boldsymbol{\cdot\nabla}\omega=-\alpha\omega+\nu\Delta\omega+\sqrt{\sigma}\eta,\,\mathbf{v}=\mathbf{e}_{z}\times\boldsymbol{\nabla}\psi,\,\omega=\Delta\psi,\label{Eq:advection}\end{equation}
} where $\omega$, $\mathbf{v}$ and $\psi$ are respectively the
vorticity, velocity and stream function ; $\alpha$ is the Rayleigh
friction coefficient and $\nu$ the viscosity. The force curl is $\eta=\Sum_{\mathbf{k}}f_{\mathbf{k}}\eta_{\mathbf{k}}\left(t\right)\exp(i\mathbf{k}.\mathbf{x})/\left(2\pi\right)$,
with $\left\{ \eta_{\mathbf{k}}\right\} $ independent Gaussian white
noises : $\left\langle \eta_{\mathbf{k}}\left(t\right)\eta_{\mathbf{k}'}\left(t'\right)\right\rangle =\delta_{\mathbf{\mathbf{k}k}'}\delta(t-t')$.
We impose $B_{0}\equiv\Sum_{\mathbf{k}}\left|f_{\mathbf{k}}\right|^{2}/\left|\mathbf{k}\right|^{2}=1$
so that $\sigma$ is the average energy injection rate.

Euler Eq. ($\alpha=\nu=\sigma=0)$ conserve the kinetic energy $E$
and vorticity moments $\Omega_{n}$ ($\Omega_{2}$ is the enstrophy)
\begin{equation}
E=\frac{1}{2}\int_{\mathcal{D}}d^{2}x\,\mathbf{v}{}^{2}\,\,\,\mbox{and}\,\,\,\Omega_{n}=\int_{\mathcal{D}}d^{2}x\,\omega^{n}.\label{Eq:energie}\end{equation}
 Application of Ito formula to the energy, and averaging over the
noise, leads to $d\left\langle E\right\rangle /dt=-2\alpha\left\langle E\right\rangle +\sigma-\nu\left\langle \Omega_{2}\right\rangle $.
If $\left\langle .\right\rangle _{S}$ denotes averages over the stationary
measure, we have $2\alpha\left\langle E\right\rangle _{S}+\nu\left\langle \Omega_{2}\right\rangle _{S}=\sigma$.
It expresses the balance between energy injection and energy dissipation.
Clearly, for flows with energetic large scales, Rayleigh friction
dominates dissipation $2\alpha\left\langle E\right\rangle _{S}\gg\nu\left\langle \Omega_{2}\right\rangle _{S}$
(mathematically we consider the limit $\nu\rightarrow0$ for fixed
$\alpha$ and assume $\nu\left\langle \Omega_{2}\right\rangle _{S}\rightarrow0$).
It is natural to fix the average energy to be of order $1$ by using
a typical turnover time as a new time unit. Putting $t'=\sqrt{\sigma/(2\alpha)}t$,
$\omega'=\sqrt{2\alpha/\sigma}\omega$, $\alpha'$=$\left(2\alpha\right)^{3/2}/(2\sigma^{1/2})$
and $\nu'=\nu\left(2\alpha/\sigma\right)^{1/2}$ and dropping the
primes, the dimensionless Eq. are {\small \begin{equation}
\frac{\partial\omega}{\partial t}+\mathbf{v}\boldsymbol{\cdot\nabla}\omega=-\alpha\omega+\nu\Delta\omega+\sqrt{2\alpha}\eta.\label{eq:NSS-rescaled}\end{equation}
} The energy balance now reads $\left\langle E\right\rangle _{S}+\left(\nu/2\alpha\right)\left\langle \Omega_{2}\right\rangle _{S}=1$.
In these dimensionless unit the Reynolds number is $1/\nu$ and the
Rayleigh number is $R_{\alpha}=\mathcal{O}\left(\mathbf{v}\boldsymbol{\cdot\nabla}\omega/\alpha\omega\right)=1/\alpha$
(arresting the inverse cascade before energy reaches the largest scale
would requires $\alpha>1$). For most geophysical flows and experiments
the case of weak forcing and dissipation is the most relevant one.
We thus study the inertial limit $\alpha\ll1$ (more precisely the
limit $\lim_{\alpha\rightarrow0}\lim_{\nu\rightarrow0}$). 

Without Rayleigh friction ($\alpha=0$), the previous discussion is
meaningless and the balance relation becomes $2\nu\left\langle \Omega_{2}\right\rangle _{S}=\sigma$.
By a natural time unit change, we can fix $\left\langle \Omega_{2}\right\rangle _{S}=1$.
The nondimensional equation is then {\small \begin{equation}
\frac{\partial\omega}{\partial t}+\mathbf{v}\boldsymbol{\cdot\nabla}\omega=\nu\Delta\omega+\sqrt{2\nu}\eta.\label{eq:NSS-viscosite}\end{equation}
}{\small \par}

From a physical point of view, this last model is less relevant than
(\ref{eq:NSS-rescaled}) but is still very interesting from an academic
point of view. A series of recent works has proved the existence of
invariant measures, validity of the law of large numbers, central
limit theorems, ergodicity and some properties of stationary measures
in the inertial limit $\nu\rightarrow0$, balance relations (see \cite{Kuksin_2004_JStatPhys_EulerianLimit}
and references therein). All following considerations are relevant
for both models (\ref{eq:NSS-rescaled},\ref{eq:NSS-viscosite}),
in their respective inertial limits.

We know since decades from real \cite{Marteau_Cardoso_Tabeling_1995PhRvE}
or numerical \cite{Schneider_Farge_2008PhysicaD,Maassen_Clercx_VanHeijst_2003JFM}
experiments, that for times large compared to the turnover time but
small compared to the dissipation time, the largest scales of 2D Navier-Stokes
turbulent flows converge towards steady solutions of Euler Eq. : \begin{equation}
\mathbf{v}\boldsymbol{\cdot\nabla}\omega=0\,\,\,\mbox{or equivalently }\,\,\,\omega=f\left(\psi\right).\label{eq:Steady_Euler}\end{equation}
 It appears to be true as well for the Euler Eq. For instance, RSM
theory predicts $f$ from given initial conditions. Given this empirical
evidence, it is thus extremely natural to expect that in the inertial
limit, measures for SNS are concentrated near steady Euler flows.
We show numerical evidences of this fact in the following.

The ensemble of steady Euler flows is huge, as it is parametrized
by the function $f$. It will be proven that when either $f$ or the
domain shape is changed, bifurcations may occur. Such abrupt transitions
lead to strong qualitative changes in the flow topology. In this critical
regime and under the action of a small random force in SNS, the system
switches randomly from one type of topology to another. In the following,
we show that this scenario is valid.

We study a bifurcation diagram for stable steady Euler solutions,
by considering \begin{equation}
S(E)=\sup_{\omega}\{{\cal S}[\omega]=\int_{\mathcal{D}}~d^{2}x~s(\omega)~|~\mathcal{E}(\omega)=E\},\label{SE}\end{equation}
 where $S(E)$ is the equilibrium entropy, ${\cal S}$ the entropy
of $\omega$ ; the specific entropy $s(\omega)=-\omega^{2}/2+\sum{\displaystyle _{n\geq2}}a_{2n}/2^{n}\omega^{2n}$
is concave assuming $s$ even for simplicity. Critical points of (\ref{SE})
verify $\omega=f(\psi)=\left(s'\right)^{-1}(-\beta\psi)$, where $\beta$
is the Lagrange multiplier associated with energy conservation. They
are thus steady Euler flows, satisfying (\ref{eq:Steady_Euler}),
and the knowledge of $f$ or $s$ are equivalent. From Arnold's theorems
\cite{Arnold_1966} or its generalization, maxima of (\ref{SE}) are
dynamically stable. One can also prove that any solutions for (\ref{SE})
are RSM equilibria \cite{Bouchet:2007_condmat}. Even if it seems
appealing, there are no clear theoretical arguments for giving a thermodynamical
interpretation to (\ref{SE}) in the SNS out-of-equilibrium context.
We thus consider (\ref{SE}) only as a practical way to describe ensembles
of \emph{stable} steady Euler solutions.

Dipoles and unidirectional flows ({}``bars'') have been obtained
numerically \cite{Yin_Montgomery_Clercx_2003PhFluids} as entropy
maxima for 2D Euler Eq. with periodic boundary conditions, assuming
$\sinh$, $\tanh$ and 3-level Poisson $\omega-\psi$ (\ref{eq:Steady_Euler})
relations. According to \cite{Yin_Montgomery_Clercx_2003PhFluids}
{}``which has the greater entropy (between dipoles and bars) depends
on seemingly arbitrary choices''.

The fact that both unidirectional flows and dipole may be equilibria
can be understood from the small energy limit of (\ref{SE}). Let
us call $\{ e_{i}\}_{i\geq1}$ the orthonormal family of eigenfunctions
of the Laplacian $-\Delta e_{i}=\lambda_{i}e_{i},~\left\langle e_{i}e_{j}\right\rangle _{\mathcal{D}}=\delta_{ij}$
($<.>_{\mathcal{D}}\equiv\int_{\mathcal{D}}~dx$ and $\lambda_{i}$
are arranged in increasing order). We decompose the vorticity as $\omega=\sum_{i\geq1}\omega_{i}e_{i}$.
The energy is then $2{\cal E}(\omega)=\sum_{i\geq1}\lambda_{i}^{-1}\omega_{i}^{2}$.
Since ${\cal E}(\omega)$ is always positive $\left\langle \omega^{2}\right\rangle _{\mathcal{D}}$
is small in the limit $E\to0$, and only the quadratic part of ${\cal S}\left[\omega\right]$
is relevant. Long but straightforward computation of (\ref{SE}) in the limit
$E\to0$ gives {\footnotesize \begin{equation}
\omega\underset{E\rightarrow0}{\sim}\left(2\lambda_{1}E\right)^{1/2}e_{1}\,\,\mbox{with}\,\, S(E)=-\lambda_{1}E+\mathcal{O}\left(a_{4}\lambda_{1}^{2}E^{2}\right).\label{eq:Entropie_Quadratique}\end{equation}
} We thus conclude that the eigenmode with the smallest eigenvalue
is selected, corresponding to the heuristic idea that energy condensate
to the largest scale. For instance when the aspect ratio $\delta>1$,
the mode $e_{1}=n_{1}\sin\left[(x+\phi_{1})/\delta\right]$ is selected.
This corresponds to a unidirectional velocity field $\mathbf{v}_{1}=n_{1}\cos\left[(x+\phi_{1})/\delta\right]\mathbf{e}_{y}$
where $\phi_{1}$ is a phase associated to the translational invariance.
A dipole is actually a mixed state $\alpha e_{1}+\beta e_{2}$ with
$e_{2}=n_{2}\sin(y+\phi_{2})$. In the weak energy limit, it can be
selected only for the degenerate case $\lambda_{1}=\lambda_{2}$.
This happens for the square box $\delta=1$. In such a case we can
prove that the degeneracy is removed by the contribution of higher
order terms in (\ref{eq:Entropie_Quadratique}). From (\ref{eq:Entropie_Quadratique}),
we conclude that the domain shape ($e_{1}$) selects the equilibria
for $\lambda_{2}-\lambda_{1}\gg a_{4}\lambda_{1}^{2}E$ whereas for
$\lambda_{2}-\lambda_{1}\ll a_{4}\lambda_{1}^{2}E$ the degeneracy
is removed by the nonlinearity of $f$ ($a_{4}$). In order to study
the bifurcation between these two behaviors, we define $g$ by $\lambda_{2}-\lambda_{1}=gE$
and we study the small energy limit $E\to0,$ with fixed $g$. Straightforward
computations lead to

\begin{equation}
S(E)=-\lambda_{1}E+E^{2}\max_{0\leq X\leq1}h(X),\label{SE5}\end{equation}
 with $h(X)=\left\langle e_{1}^{4}\right\rangle _{\mathcal{D}}a_{4}\lambda_{1}^{2}-gX+2\gamma\lambda_{1}^{2}a_{4}X(1-X)$,
where $\gamma=3\left\langle e_{1}^{2}e_{2}^{2}\right\rangle _{\mathcal{D}}-\left\langle e_{1}^{4}\right\rangle _{\mathcal{D}}>0$.
The vorticity equilibria is then $\omega_{eq}\sim_{E\rightarrow0}\sqrt{2\lambda_{1}E(1-X_{M})}e_{1}+\sqrt{2E\lambda_{1}X_{M}}e_{2}$
where $X_{M}$ is the maximizer of $h$ in (\ref{SE5}). For $X_{M}=0$
or $X_{M}=1$, $\omega_{eq}$ is an unidirectional flow whereas for
$0<X_{M}<1$ it is a dipole (symmetric for $X_{M}=1/2$). The selection
occurs via maximization of $h$. When maximizing $h$, the sign of
the parameter $a_{4}$ plays a crucial role. We note that $a_{4}$
is intimately related to the shape of the relationship $\omega=f(\psi)=\left(s'\right)^{-1}(-\beta\psi)$.
Indeed $\left(s'\right)^{-1}(-x)=x+a_{4}x^{3}+o(x^{3})$ and when
$a_{4}>0$ (resp. $a_{4}<0$), the curve $f(\psi)$ bends upward (resp.
downward) for positive $\psi$ similarly to $\sinh$ (resp. $\tanh$).

\begin{figure}
\includegraphics[width=0.5\columnwidth,keepaspectratio]{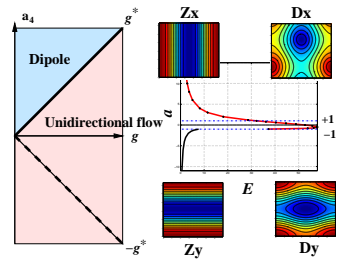}

\caption{\label{fig:Equilibre} Bifurcation diagrams for steady Euler flows
a) in the $g$-$a_{4}$ plane b) obtained numerically in an $E-a_{4}$
plane with $\delta=1.1$. The inset illustrates good agreement between
numerical and theoretical results in the low energy limit.}
\end{figure}

The bifurcation diagram is summarized in Fig. \ref{fig:Equilibre}
a). In the degenerate case ($g=0$), the dipole is selected for $a_{4}>0$
($\sinh$ like), whereas unidirectional flows are selected for $a_{4}<0$
($\tanh$ like). The term $-gX$ favors the pure state $e_{1}$ ($X=0$).
For $a_{4}<0$ the unidirectional flow $e_{1}$ is always selected.
More interestingly, for $a_{4}>0$ a bifurcation occurs along the
critical line $g^{\star}=2\gamma\lambda_{1}^{2}a_{4}$ between dipole
and unidirectional flows. %

We have obtained the bifurcation diagram in the limit of small energy
using the scaling $\lambda_{2}-\lambda_{1}=gE$. From a practical
point of view, it is more convenient to work for a fixed aspect ratio
$\delta$. Using the relation $g=(\lambda_{2}-\lambda_{1})/E$, we
obtain that the critical line in a $E-a_{4}$ plane is the hyperbola
$a_{4}E=8\pi^{2}\left(\delta-1\right)/3+o(\delta-1)$. We use a continuation
algorithm in order to numerically compute solution to (\ref{SE})
corresponding to $f_{a_{4}}(x)=\left(1/3-2a_{4}\right)\tanh x+\left(2/3+2a_{4}\right)\sinh x$.
The inset of Fig \ref{fig:Equilibre} \textbf{b)} shows good agreement
for transition lines obtained either with the continuation algorithm
or the low-energy limit theoretical result, for $\delta=1.01$. Figure
\ref{fig:Equilibre} \textbf{b)} shows the bifurcation diagram for
$\delta=1.1$; in such a case the transition line is still very close
to an hyperbola provided energy is small. 

Following the same reasoning, small-energy bifurcation diagrams could
be computed for any Euler-like model like QG or SW models. Most often,
the domain shape selects the flow topology. When domain shape is varied,
we meet eigenvalues degeneracy. In all these cases a bifurcation diagram
can be computed where the transition line corresponds to the competition
between the $\omega-\psi$ (\ref{eq:Steady_Euler}) nonlinearity and
the domain shape.\\

\begin{figure}[htpb]
\includegraphics[width=0.8\columnwidth,keepaspectratio]{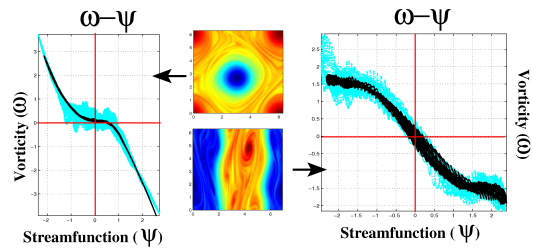}

\caption{\label{fig:Omega-Psi} $\omega-\psi$ scatter-plots (cyan). In black
the same after time averaging (averaging windows $1\ll\tau\ll1/\nu$,
the drift due to translational invariance has been removed) a) dipole
case with $\delta=1.03$ b) unidirectional case $\delta=1.10$. }
\end{figure}

We expect to observe both dipoles and unidirectional flows in SNS.
Numerical simulations in a square domain $\delta=1$ exhibit statistically
stationary $\omega$ with a dipole structure (Fig. \ref{fig:Omega-Psi}
\textbf{a)}), whereas for $\delta\geq1.1$, nearly unidirectional
flows are observed (Fig. \ref{fig:Omega-Psi} \textbf{b)}). This result
has been confirmed both for $\alpha=0$ and $\alpha\neq0$, and for
different $\nu$ values and forcing spectra. One observes in Fig.
\ref{fig:Omega-Psi} a $\omega-\psi$ relation qualitatively similar
to a sinh, in the dipole case and to a tanh in the unidirectional
case. This confirms that $\omega$ remains close to steady Euler flows.

A very natural order parameter is $\left|z_{1}\right|$, where $z_{1}=\frac{1}{\left(2\pi\right)^{2}}\left\langle \omega(x,y)\exp(iy)\right\rangle _{\mathcal{D}}$.
Indeed, for unidirectional flow $\omega=\alpha e_{1}$, $z_{1}=0$,
whereas for a dipole $\omega=\alpha\left(e_{1}+e_{2}\right)$, $\left|z_{1}\right|=\alpha$.
Fig. \ref{ts_f} shows $\left|z_{1}\right|$ time series for $\delta=1.02$
and $\delta=1.04$. The remarkable observation is the bimodal behavior
in this transition range. The switches from $\left|z_{1}\right|$
values close to zero to values of order of $0.6$ correspond to genuine
transitions between unidirectional and dipole flows. The PDF of the
complex variable $z_{1}$ (Fig. \ref{ts_f}) exhibits a circle corresponding
to the dipole state (a slow dipole random translation corresponds
into to a phase drift for $z_{1}$, explaining the circular symetry).
The zonal state corresponds to the central peak. As $\delta$ increases,
one observes less occurrences of the dipole. For larger (resp. smaller)
values of $\delta$ only unidirectional (dipole) flows exist. The
transition is also visible in other physical variables. For instance
$\Omega_{4}=\left\langle |\omega|^{4}\right\rangle _{\mathcal{D}}$
switches between a state with weak variance and low mean value (unidirectional)
to an intermittent state with large variance and larger mean value
(dipole). Topology changes are very slow dynamical processes : for
the model (\ref{eq:NSS-viscosite}), an average transition time is
of order $1/\nu$. For instance Fig. \ref{ts_f} represents $3.10^{4}$
turnover times. For this reason, because of numerical limitations
it has not been yet possible to obtain convincing analysis of the
switch time statistics.

In the spirit of \cite{Benzi_2005PRL} we look for low-dimensional
analogies. When looking how evolves the PDF for the order parameter,
while changing the control parameter, the dipole-unidirectional transition
has striking similarities with the stochastic differential equation
\begin{equation}
dx=x(\mu+x^{2}-x^{4})dt+\sigma dW.\label{sdebi}\end{equation}
 The deterministic part of (\ref{sdebi}) is the normal form for a
generalized subcritical pitchfork bifurcation. For $\mu<-1/4$, one
has a single stable fixed point $x^{\star}=0$. For $\mu>0$ there
are three fixed point, one unstable $x_{0}=0$ and two stable $x_{1,2}=\pm(1+(1+4\mu)^{1/2})^{1/2}$.
For $\mu\in]-1/4,0[$, three stable fixed points coexist ($x_{0}$
and $x_{1,2}$) and two unstable ones $x_{3,4}=\pm(1-(1+4\mu)^{1/2})^{1/2}$.
With additive noise $(\sigma\neq0)$, when $\mu<-1/4$, the $\left|x\right|$
PDF has a single peak centered at $x=0$. In the interval corresponding
to $\mu\in]-1/4,0[$, an additional peak appears related to $\left|x_{1,2}\right|$.
Finally, there is a transition for $\mu$ larger than 0 and only one
peak corresponding to $\left|x_{1}\right|$ remains. 

However, we stress that a low-dimensional model like (\ref{sdebi}),
as useful as it may be, lacks part of the phenomena. For instance,
it can not explain why $\Omega_{4}$ is intermittent while $\left|z_{1}\right|$
is not. Moreover, the role of turbulence here is not only to act as
noise, but also to build up the large-scale flow by inverse cascade.
The inverse cascade properties are strongly affected by the existing
large-scale flow, leading to the observed self-organization process.
From a theoretical point of view, the main issue, beyond the scope
of this letter, is to explain which of the Euler steady states will
be selected by turbulence and to predict the relative frequency of
such states. We thus need an alternative theoretical approach bridging
the gap between the two classical theories : self-similar inverse
energy cascade on one hand and RSM equilibrium statistical mechanics
on the other hand. 

In this letter, we have not addressed the other crucial issue : fluctuations.
Some very interesting results and considerations on small-scale fluctuations
for turbulence dominated by large-scale flows may be found in \cite{Dubrulle_Nazarenko_1997PhyD,Nazarenko_Laval_JFM_2000,Chertkov_Connaughton_andco_2007_PRL_EnergyCondesation}.
In forthcoming works, the statistical properties of these random change
of flow topologies and of fluctuations will be investigated. Finally,
it will also be extremely interesting to analyze the connexions with
similar transitions observed in other contexts \cite{Sommeria_1986_JFM_2Dinverscascade_MHD,Berhanu_etc_Fauve_2007_EPL_MagneticFieldReversal,Weeks_Tian_etc_Swinney_Ghil_Science_1997,Schmeits_Dijkstraa_2001_JPO_BimodaliteGulfStream}.

We infer similar random flow topology changes for other geometry for
2D SNS, QG and SW models. Using simple generalization of our analysis,
rotating tanks experiments can be designed in order to observe similar
phenomena. This study also suggests that flows like the Kuroshio currents
\cite{Schmeits_Dijkstraa_2001_JPO_BimodaliteGulfStream} or the Gulf
Stream might be close to steady solutions of inertial models.

This work was supported by ANR program STATFLOW (ANR-06-JCJC-0037-01).

\bibliographystyle{apsrev} \bibliographystyle{apsrev}
\bibliography{FBouchet,Long_Range,Meca_Stat_Euler,Experimental_2D_Flows,Stochastic_Processes,NS-Stochastic,Turbulence_2D,Euler_Stability,Ocean}

\begin{thebibliography}{19}
\expandafter\ifx\csname natexlab\endcsname\relax\def\natexlab#1{#1}\fi
\expandafter\ifx\csname bibnamefont\endcsname\relax
  \def\bibnamefont#1{#1}\fi
\expandafter\ifx\csname bibfnamefont\endcsname\relax
  \def\bibfnamefont#1{#1}\fi
\expandafter\ifx\csname citenamefont\endcsname\relax
  \def\citenamefont#1{#1}\fi
\expandafter\ifx\csname url\endcsname\relax
  \def\url#1{\texttt{#1}}\fi
\expandafter\ifx\csname urlprefix\endcsname\relax\def\urlprefix{URL }\fi
\providecommand{\bibinfo}[2]{#2}
\providecommand{\eprint}[2][]{\url{#2}}

\bibitem[{\citenamefont{{Berhanu} et~al.}(2007)\citenamefont{{Berhanu},
  {Monchaux}, {Fauve}, {Mordant}, {Petrelis}, {Chiffaudel}, {Daviaud},
  {Dubrulle}, {Marie}, {Ravelet}
  et~al.}}]{Berhanu_etc_Fauve_2007_EPL_MagneticFieldReversal}
\bibinfo{author}{\bibfnamefont{M.}~\bibnamefont{{Berhanu}}},
  \bibinfo{author}{\bibfnamefont{R.}~\bibnamefont{{Monchaux}}},
  \bibinfo{author}{\bibfnamefont{S.}~\bibnamefont{{Fauve}}},
  \bibinfo{author}{\bibfnamefont{N.}~\bibnamefont{{Mordant}}},
  \bibinfo{author}{\bibfnamefont{F.}~\bibnamefont{{Petrelis}}},
  \bibinfo{author}{\bibfnamefont{A.}~\bibnamefont{{Chiffaudel}}},
  \bibinfo{author}{\bibfnamefont{F.}~\bibnamefont{{Daviaud}}},
  \bibinfo{author}{\bibfnamefont{B.}~\bibnamefont{{Dubrulle}}},
  \bibinfo{author}{\bibfnamefont{L.}~\bibnamefont{{Marie}}},
  \bibinfo{author}{\bibfnamefont{F.}~\bibnamefont{{Ravelet}}},
  \bibnamefont{et~al.}, \bibinfo{journal}{Eur. Phys. Lett.}
  (\bibinfo{year}{2007}).

\bibitem[{\citenamefont{Ravelet et~al.}(2004)\citenamefont{Ravelet, Mari\'e,
  Chiffaudel, and Daviaud}}]{Ravelet_Marie_Chiffaudel_Daviaud_PRL2004}
\bibinfo{author}{\bibfnamefont{F.}~\bibnamefont{Ravelet}},
  \bibinfo{author}{\bibfnamefont{L.}~\bibnamefont{Mari\'e}},
  \bibinfo{author}{\bibfnamefont{A.}~\bibnamefont{Chiffaudel}},
  \bibnamefont{and} \bibinfo{author}{\bibfnamefont{F.~m.~c.}
  \bibnamefont{Daviaud}}, \bibinfo{journal}{Phys. Rev. Lett.}
  \textbf{\bibinfo{volume}{93}}, \bibinfo{pages}{164501}
  (\bibinfo{year}{2004}).

\bibitem[{\citenamefont{{Weeks} et~al.}(1997)\citenamefont{{Weeks}, {Tian},
  {Urbach}, {Ide}, {Swinney}, and
  {Ghil}}}]{Weeks_Tian_etc_Swinney_Ghil_Science_1997}
\bibinfo{author}{\bibfnamefont{E.~R.} \bibnamefont{{Weeks}}},
  \bibinfo{author}{\bibfnamefont{Y.}~\bibnamefont{{Tian}}},
  \bibinfo{author}{\bibfnamefont{J.~S.} \bibnamefont{{Urbach}}},
  \bibinfo{author}{\bibfnamefont{K.}~\bibnamefont{{Ide}}},
  \bibinfo{author}{\bibfnamefont{H.~L.} \bibnamefont{{Swinney}}},
  \bibnamefont{and} \bibinfo{author}{\bibfnamefont{M.}~\bibnamefont{{Ghil}}},
  \bibinfo{journal}{Science} \textbf{\bibinfo{volume}{278}},
  \bibinfo{pages}{1598} (\bibinfo{year}{1997}).

\bibitem[{\citenamefont{{Sommeria}}(1986)}]{Sommeria_1986_JFM_2Dinverscascade_%
MHD}
\bibinfo{author}{\bibfnamefont{J.}~\bibnamefont{{Sommeria}}},
  \bibinfo{journal}{J. Fluid. Mech.} \textbf{\bibinfo{volume}{170}},
  \bibinfo{pages}{139} (\bibinfo{year}{1986}).

\bibitem[{\citenamefont{{Maassen} et~al.}(2003)\citenamefont{{Maassen},
  {Clercx}, and {van Heijst}}}]{Maassen_Clercx_VanHeijst_2003JFM}
\bibinfo{author}{\bibfnamefont{S.~R.} \bibnamefont{{Maassen}}},
  \bibinfo{author}{\bibfnamefont{H.~J.~H.} \bibnamefont{{Clercx}}},
  \bibnamefont{and} \bibinfo{author}{\bibfnamefont{G.~J.~F.} \bibnamefont{{van
  Heijst}}}, \bibinfo{journal}{J. Fluid Mech.} \textbf{\bibinfo{volume}{495}},
  \bibinfo{pages}{19} (\bibinfo{year}{2003}).

\bibitem[{\citenamefont{Schmeits and
  Dijkstraa}(2001)}]{Schmeits_Dijkstraa_2001_JPO_BimodaliteGulfStream}
\bibinfo{author}{\bibfnamefont{M.~J.} \bibnamefont{Schmeits}} \bibnamefont{and}
  \bibinfo{author}{\bibfnamefont{H.~A.} \bibnamefont{Dijkstraa}},
  \bibinfo{journal}{J. Phys. Oceanogr.} \textbf{\bibinfo{volume}{31}},
  \bibinfo{pages}{3425} (\bibinfo{year}{2001}).

\bibitem[{\citenamefont{{Benzi}}(2005)}]{Benzi_2005PRL}
\bibinfo{author}{\bibfnamefont{R.}~\bibnamefont{{Benzi}}},
  \bibinfo{journal}{Phys. Rev. Lett.} \textbf{\bibinfo{volume}{95}},
  \bibinfo{pages}{024502} (\bibinfo{year}{2005}), \eprint{arXiv:nlin/0410048}.

\bibitem[{\citenamefont{{Kraichnan}}(1967)}]{Kraichnan_Phys_Fluid_1967_2Dturbu%
lence}
\bibinfo{author}{\bibfnamefont{R.~H.} \bibnamefont{{Kraichnan}}},
  \bibinfo{journal}{Phys. Fluids} \textbf{\bibinfo{volume}{10}},
  \bibinfo{pages}{1417} (\bibinfo{year}{1967}).

\bibitem[{\citenamefont{{Bernard} et~al.}(2006)\citenamefont{{Bernard},
  {Boffetta}, {Celani}, and
  {Falkovich}}}]{Bernard_Boffetta_Celani_Falkovich_2006Nature}
\bibinfo{author}{\bibfnamefont{D.}~\bibnamefont{{Bernard}}},
  \bibinfo{author}{\bibfnamefont{G.}~\bibnamefont{{Boffetta}}},
  \bibinfo{author}{\bibfnamefont{A.}~\bibnamefont{{Celani}}}, \bibnamefont{and}
  \bibinfo{author}{\bibfnamefont{G.}~\bibnamefont{{Falkovich}}},
  \bibinfo{journal}{Nature Physics} \textbf{\bibinfo{volume}{2}},
  \bibinfo{pages}{124} (\bibinfo{year}{2006}), \eprint{arXiv:nlin/0602017}.

\bibitem[{\citenamefont{{Eyink} and
  {Sreenivasan}}(2006)}]{Eyink_Sreenivasan_2006_Rev_Modern_Physics}
\bibinfo{author}{\bibfnamefont{G.~L.} \bibnamefont{{Eyink}}} \bibnamefont{and}
  \bibinfo{author}{\bibfnamefont{K.~R.} \bibnamefont{{Sreenivasan}}},
  \bibinfo{journal}{Rev. Mod. Phys.} \textbf{\bibinfo{volume}{78}},
  \bibinfo{pages}{87} (\bibinfo{year}{2006}).

\bibitem[{\citenamefont{{Chertkov} et~al.}(2007)\citenamefont{{Chertkov},
  {Connaughton}, {Kolokolov}, and
  {Lebedev}}}]{Chertkov_Connaughton_andco_2007_PRL_EnergyCondesation}
\bibinfo{author}{\bibfnamefont{M.}~\bibnamefont{{Chertkov}}},
  \bibinfo{author}{\bibfnamefont{C.}~\bibnamefont{{Connaughton}}},
  \bibinfo{author}{\bibfnamefont{I.}~\bibnamefont{{Kolokolov}}},
  \bibnamefont{and}
  \bibinfo{author}{\bibfnamefont{I.}~\bibnamefont{{Lebedev}}},
  \bibinfo{journal}{Phys. Rev. Lett.} \textbf{\bibinfo{volume}{99}},
  \bibinfo{pages}{084501} (\bibinfo{year}{2007}).

\bibitem[{\citenamefont{{Kuksin}}(2004)}]{Kuksin_2004_JStatPhys_EulerianLimit}
\bibinfo{author}{\bibfnamefont{S.~B.} \bibnamefont{{Kuksin}}},
  \bibinfo{journal}{J. Stat. Phys.} \textbf{\bibinfo{volume}{115}},
  \bibinfo{pages}{469} (\bibinfo{year}{2004}).

\bibitem[{\citenamefont{{Marteau} et~al.}(1995)\citenamefont{{Marteau},
  {Cardoso}, and {Tabeling}}}]{Marteau_Cardoso_Tabeling_1995PhRvE}
\bibinfo{author}{\bibfnamefont{D.}~\bibnamefont{{Marteau}}},
  \bibinfo{author}{\bibfnamefont{O.}~\bibnamefont{{Cardoso}}},
  \bibnamefont{and}
  \bibinfo{author}{\bibfnamefont{P.}~\bibnamefont{{Tabeling}}},
  \bibinfo{journal}{\pre} \textbf{\bibinfo{volume}{51}}, \bibinfo{pages}{5124}
  (\bibinfo{year}{1995}).

\bibitem[{\citenamefont{{Schneider} and
  {Farge}}(2008)}]{Schneider_Farge_2008PhysicaD}
\bibinfo{author}{\bibfnamefont{K.}~\bibnamefont{{Schneider}}} \bibnamefont{and}
  \bibinfo{author}{\bibfnamefont{M.}~\bibnamefont{{Farge}}},
  \bibinfo{journal}{Physica D}  (\bibinfo{year}{2008}).

\bibitem[{\citenamefont{{Arnold}}(1966)}]{Arnold_1966}
\bibinfo{author}{\bibfnamefont{V.~I.} \bibnamefont{{Arnold}}},
  \bibinfo{journal}{Izv. Vyssh. Uchebbn. Zaved. Matematika; Engl. transl.: Am.
  Math. Soc. Trans.} \textbf{\bibinfo{volume}{79}}, \bibinfo{pages}{267}
  (\bibinfo{year}{1966}).

\bibitem[{\citenamefont{{Bouchet}}(2007)}]{Bouchet:2007_condmat}
\bibinfo{author}{\bibfnamefont{F.}~\bibnamefont{{Bouchet}}},
  \bibinfo{journal}{ArXiv e-prints}  (\bibinfo{year}{2007}),
  \eprint{0710.5094}.

\bibitem[{\citenamefont{{Yin} et~al.}(2003)\citenamefont{{Yin}, {Montgomery},
  and {Clercx}}}]{Yin_Montgomery_Clercx_2003PhFluids}
\bibinfo{author}{\bibfnamefont{Z.}~\bibnamefont{{Yin}}},
  \bibinfo{author}{\bibfnamefont{D.~C.} \bibnamefont{{Montgomery}}},
  \bibnamefont{and} \bibinfo{author}{\bibfnamefont{H.~J.~H.}
  \bibnamefont{{Clercx}}}, \bibinfo{journal}{Phys. Fluids}
  \textbf{\bibinfo{volume}{15}}, \bibinfo{pages}{1937} (\bibinfo{year}{2003}),
  \eprint{arXiv:physics/0211024}.

\bibitem[{\citenamefont{{Dubrulle} and
  {Nazarenko}}(1997)}]{Dubrulle_Nazarenko_1997PhyD}
\bibinfo{author}{\bibfnamefont{B.}~\bibnamefont{{Dubrulle}}} \bibnamefont{and}
  \bibinfo{author}{\bibfnamefont{S.}~\bibnamefont{{Nazarenko}}},
  \bibinfo{journal}{Physica D} \textbf{\bibinfo{volume}{110}},
  \bibinfo{pages}{123} (\bibinfo{year}{1997}).

\bibitem[{\citenamefont{{Nazarenko} and
  {Laval}}(2000)}]{Nazarenko_Laval_JFM_2000}
\bibinfo{author}{\bibfnamefont{S.}~\bibnamefont{{Nazarenko}}} \bibnamefont{and}
  \bibinfo{author}{\bibfnamefont{J.-P.} \bibnamefont{{Laval}}},
  \bibinfo{journal}{J. Fluid Mech.} \textbf{\bibinfo{volume}{408}},
  \bibinfo{pages}{301} (\bibinfo{year}{2000}).

\end{thebibliography}

\end{document}